\documentclass[twocolumn,showpacs,preprintnumbers,amsmath,amssymb,superscriptaddress]{revtex4}
\usepackage{graphicx}
\usepackage{bm}

\usepackage[usenames]{color}

\begin{document}


\title{
Topological and Transport Properties of Dirac Fermions 
in Antiferromagnetic Metallic Phase 
of Iron-Based Superconductors}


\author{Takao Morinari}\email{morinari@yukawa.kyoto-u.ac.jp}
\affiliation{Yukawa Institute for Theoretical Physics, Kyoto University, Kyoto 606-8502, Japan}
\author{Eiji Kaneshita}
\affiliation{Yukawa Institute for Theoretical Physics, Kyoto University, Kyoto 606-8502, Japan}
\author{Takami Tohyama}
\affiliation{Yukawa Institute for Theoretical Physics, Kyoto University, Kyoto 606-8502, Japan}
\affiliation{JST, Transformative Research-Project on Iron Pnictides (TRIP), Chiyoda, Tokyo 102-0075, Japan}

\date{\today}

\begin{abstract}
We investigate Dirac fermions in the antifferomagnetic metallic 
state of iron-based superconductors. 
Deriving an effective Hamiltonian for
Dirac fermions,
we reveal that there exist two Dirac cones carrying the same chirality, 
contrary to graphene,
compensated by a Fermi surface with a quadratic energy dispersion
as a consequence of a non-trivial topological property 
inherent in the band structure.
We also find that the presence of the Dirac fermions
gives the difference of sign-change temperatures 
between the Hall coefficient and the thermopower. 
This is consistent with available experimental data.
\end{abstract}


\maketitle

\newcommand{\be}{\begin{equation}}
\newcommand{\ee}{\end{equation}}
\newcommand{\bea}{\begin{eqnarray}}
\newcommand{\eea}{\end{eqnarray}}

The iron-pnictide high-temperature superconductors have attracted
much attention since the discovery \cite{Kamihara08}.
In this multi-orbital system, the Fermi surfaces consist of 
electron and hole Fermi surfaces of comparable sizes.
This multi-band structure leads to the rich physics,
presumably including superconductivity
with high-transition temperatures \cite{Mazin08,Kuroki08}.
In the parent antiferromagnetic phase,
the system remains metallic \cite{Sebastian08,Analytis09} 
contrary to the simple antiferromagnetic state in single band systems.
This parent antiferromagnetic state possesses a non-trivial
topological property; consequently appears
a Dirac fermion energy spectrum close to the Fermi energy \cite{Ran09}.
In this Letter, we explore the Dirac fermion physics in this system.

Dirac fermions in condensed matter systems were first highlighted
by the observation of the anomalous integer quantum Hall effect 
in graphene \cite{Novoselov2005,Zhang2005}.
In graphene, the energy spectrum is linear at the corners of the first Brillouin zone,
so that the electrons at low energies are well described 
by the Dirac equation with the speed of light being replaced 
by the Fermi velocity \cite{Wallace47}.
Such a unique energy band structure gives rise to several 
characteristic transport properties \cite{Geim07}.

In the iron-pnictide superconductors, a similar linear spectrum
was discussed by Ran {\it et al.}\cite{Ran09}
They pointed out that hybridizations between the Fe $3d$ orbitals 
and the pnictide ion $4p$ orbitals
give rise to the band degeneracy characterized by a non-trivial topology.
Contrary to a conventional spin density wave (SDW), 
there are gapless nodal points along the Fermi surface.
The topology here is characterized by ``vorticity" quantum number,
which is associated with the phase winding defined through
a two-component spinor wave function.
The Fermi surfaces connected by the SDW wave vector
have a vorticity zero and a vorticity two:
This vorticity mismatch leads to a nodal SDW, and creates 
Dirac cones near the Fermi energy.

In this Letter 
we shall derive the effective theory describing the Dirac fermions
based on the five-band model \cite{Kuroki08}
with the SDW mean field analysis \cite{Ran09,Kaneshita09}.
We reveal that there exist two Dirac cones carrying the {\it same} chirality
contrary to the Dirac fermions in graphene.
This unusual feature is due to the presence of another Fermi surface, 
which does not have a linear spectrum but has chirality.
We demonstrate that the scattering rate difference
between Dirac fermions and conventional electrons
leads to anomalous temperature dependence
of the transport coefficients
which are consistent with the experiment \cite{Matusiak10}.

\begin{figure}[t]
   \begin{center}
    \includegraphics[width=0.8 \linewidth]{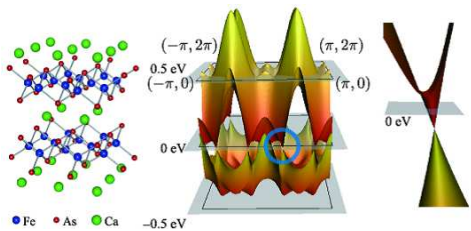}
   \end{center}
   \caption{ \label{fig_bands}
	(color online)
	(Left) The crystal structure of CaFe$_2$As$_2$.
	(Middle) 
	The electron band structure near the Fermi energy
	in the SDW state.
	The unit cell is taken by the square lattice formed by Fe atoms.
	(Right) The magnification of the band structure 
	near the Dirac point indicated by the circle in the left.
    }
 \end{figure}

For the band structure calculation, we take the five band model
\bea
\mathcal{H}_0  
&=& \sum\limits_{i,j} {\sum\limits_{\mu ,\nu } 
{\sum\limits_s  {\left( {t_{i\mu ,j\nu }  
+ \delta _{ij} \delta _{\mu \nu } } \varepsilon _\mu  \right) 
d_{i\mu s }^\dag  d_{j\nu s } } } }  
\nonumber \\ 
&=& \sum\limits_{{\mathbf{k}},\mu ,\nu ,s } 
{\varepsilon _{\mathbf{k}}^{\mu \nu } d_{{\mathbf{k}}\mu s }^\dag  
d_{{\mathbf{k}}\nu s } },
\eea
where $d_{j \mu s}^{\dagger}$ creates an electron with spin $s$
on the $\mu$-th orbital at site $j$.
The parameters $t_{i\mu ,j\nu }$ and $\varepsilon _\mu$
are given in Ref.~\cite{Kuroki08}.
Following Ref.~\cite{Ran09} we take 
the following interaction form:
\bea
\mathcal{H}_I &=& U\sum\limits_{j,\mu } {n_{j\mu  \uparrow } n_{j\mu  \downarrow } }  
+ \left( {U - 2J} \right)\sum\limits_{j,\mu  < \nu, s, s' } {n_{j\mu s} n_{j\nu s'} } \nonumber \\
& &  + J\sum\limits_{j,\mu  < \nu ,s ,s'} {d_{j\mu s }^\dag  
d_{j\nu s'}^\dag  d_{j\mu s'} d_{j\nu s} } 
\nonumber \\
& &  + J\sum\limits_{j,\mu  < \nu } {\left( {d_{j\mu  \uparrow }^\dag  
d_{j\mu  \downarrow }^\dag  d_{j\nu  \downarrow } d_{j\nu  \uparrow }  
+ {\text{H}}{\text{.c}}{\text{.}}} \right)},
\eea
where $n_{j\mu s}  = d_{j\mu s }^\dag  d_{j\mu s}^\dag$.
The mean field Hamiltonian for the SDW state reads,
\be
\mathcal{H}_{MF}  = \sum\limits_{{\mathbf{k}} \in RBZ,\mu ,\nu , s } 
{\left( {\begin{array}{*{20}c}
   {d_{{\mathbf{k}}\mu s}^\dag  } & {d_{{\mathbf{k}} + {\mathbf{Q}},\mu s}^\dag  }  \\
 \end{array} } \right)\left[ {\mathcal{H}_{{\mathbf{k}} s} } \right]_{\mu \nu } 
\left( {\begin{array}{*{20}c}
   {d_{{\mathbf{k}}\nu s} }  \\
   {d_{{\mathbf{k}} + {\mathbf{Q}},\nu s} }  \\
 \end{array} } \right)},
\ee
where ${\bf Q}=(\pi,0)$ and 
the ${\bf k}$ summation is taken over the reduced Brillouin zone.
The matrix ${\mathcal{H}_{{\mathbf{k}} s} }$ is given by
\be
\left[ {\mathcal{H}_{{\mathbf{k}} s} } \right]_{\mu \nu }  
= \left( {\begin{array}{*{20}c}
   {\varepsilon _{\mathbf{k}}^{\mu \nu } } & { \pm m_{SDW}^{\mu \nu } }  \\
   {\pm m_{SDW}^{\mu \nu } } & {\varepsilon _{{\mathbf{k}} + {\mathbf{Q}}}^{\mu \nu } }  \\
 \end{array} } \right),
\ee
where we take the plus (minus) sign for $s=\uparrow (\downarrow)$.
The mean field order parameters, $m_{SDW}^{\mu \nu}$,
are determined by solving the mean field equations in a self-consistent way.
Figure \ref{fig_bands} shows the electron band structure 
calculated in the SDW state with $U=1.2$~eV and $J=0.25$~eV
with the zero temperature magnetic moment $1.0 \mu_B$,
which is close to the value observed by neutron scattering 
in BaFe$_2$As$_2$ \cite{Huang08}.

In the SDW state, Dirac points appear near the Fermi energy.
Those Dirac points exist on the $k_x$ axis and,
in some parameter region, on the $k_y$ axis as well
but not at the high symmetric points in the Brillouin zone.
Since the Dirac points appearing on the $k_y$ axis are not protected  
by topology \cite{Ran09}, we focus on the Dirac points on the $k_x$ axis.
These Dirac points are stable as long as the magnetic moment 
is less than $3\mu_B$.

In order to investigate the 
physical
properties of the Dirac fermions
we derive the effective Hamiltonian around the Dirac point
following Ref.~\cite{Kobayashi07},
in which the Dirac fermions in the organic conductor
$\alpha$-(BEDT-TTF)$_2$I$_3$ were analyzed.
The effective Hamiltonian is used to determine 
chirality and the Landau level 
structure of Dirac fermions.

First we expand the Hamiltonian $\mathcal{H}_{\bf k}$ 
around the Dirac point ${\bf k}_c$ with denoting
${\bf k} = {\bf k}_c + \delta {\bf k}$,
\be
\mathcal{H}_{\mathbf{k}}  = \mathcal{H}_{{\mathbf{k}}_c }  
+ 
\frac{{\partial \mathcal{H}_{{\mathbf{k}}_c } }}{{\partial k_x }}
\delta k_x 
+ 
\frac{{\partial \mathcal{H}_{{\mathbf{k}}_c } }}{{\partial k_y }}
\delta k_y 
+ ...
\ee
(The spin index is suppressed hereafter.)
Next, noting that a Dirac point is formed by two eigenstates
we construct the effective $2 \times 2$ Hamiltonian
using those two eigenstates denoted by $\left| {{\bf{k}}, \pm } \right\rangle $.
The general form of the effective Hamiltonian is \cite{Kobayashi07}, 
\be
\mathcal{H}_{\delta \mathbf{k}}^{eff}  = \sum\limits_{\rho  = 0,x,y,z} 
{{\delta \mathbf{k}} 
\cdot {\mathbf{v}}_\rho  \sigma _\rho  },
\ee
where ${\bm \sigma} = (\sigma_x, \sigma_y, \sigma_z)$ are
the Pauli matrices and $\sigma_0$ is the unit matrix.
(We set $\hbar = 1$.)
The parameters ${\bf v}_{\rho}$ are determined by the following equation
with $\alpha=x,y$:
\be
\left\langle {{\bf{k}},s } \right|\frac{{\partial {\cal H}_{{\bf{k}}_c } }}
{{\partial k_{\alpha}  }}\left| {{\bf{k}},s' } \right\rangle  
= \sum\limits_{\rho  = 0,x,y,z} {v_\rho^{\alpha}
\left( {\sigma _\rho  } \right)_{ss'} }. 
\ee

\begin{figure}[t]
   \begin{center}
    \includegraphics[width=0.9 \linewidth]{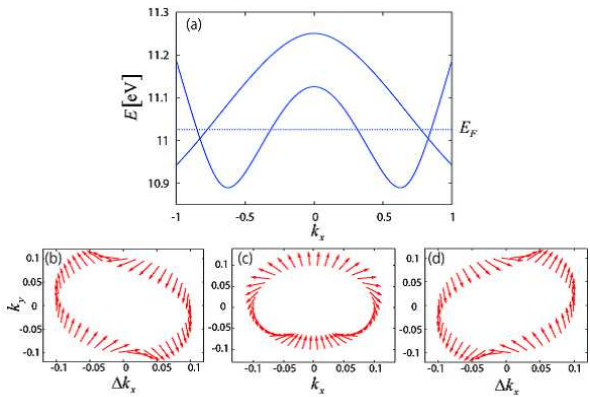}
   \end{center}
\caption{ 
(color online)
(a) The energy bands along the $k_x$ axis.
The dotted line represents the Fermi energy.
Two Dirac points are located at $(\pm 0.829,0)$.
The direction of the vector ${\bf n}({\bf k})=(n_x({\bf k}),n_z({\bf k}))$ around 
(a) $(-0.829,0)$,
(b) $(0,0)$,
and (c) $(0.829,0)$.
}
\label{fig_chirality}
\end{figure}

Figure \ref{fig_chirality}(a) shows the energy band dispersion
of the state (shown in Fig.~\ref{fig_bands})
along the $k_x$ axis.
The Dirac points are at ${\bf k}_c=(\pm 0.829,0)$.
The positions of the Dirac points are close to the values
reported in angle resolved photoemission 
spectroscopy measurements \cite{Richard09}.
The effective Hamiltonian is given by
\be
{\cal H}_{\bf k}  = \left( {E_c  \pm v_0^x k_x } \right)\sigma _0  
\pm \left( {v_z^x k_x \sigma _z  + v_x^y k_y \sigma _x } \right),
\label{eq_DiracH}
\ee
where $E_c-E_F=-20$~meV, $v_0^x/a=0.286$~eV,
$v_x^y/a = 0.229$~eV, $v_z^x/a=0.672$~eV
with $a$ the lattice constant.
The non-zero value of $v_0^x$ implies that the Dirac cone 
is tilted in the $k_x$ direction.
As a result, we expect a strong anisotropy in the Fermi velocity.

Now we discuss chirality associated with Dirac fermions.
In Fig.~\ref{fig_chirality}(b) and (d), the vector 
${\bf n}_{\bf k} \equiv \langle {\bf k},+| {\bm \sigma} | {\bf k},+\rangle$
is shown around the two Dirac points.
The vector field ${\bf n}_{\bf k}$ shows a vortex configuration
with vorticity one.
A remarkable fact is that the two Dirac cones have the same vorticity:
The vector ${\bf n}_{\bf k}$ is rotated clockwise when
one goes around each Dirac point.
In conventional Dirac fermion systems, like graphene,
two Dirac cones have the opposite chirality by symmetry.
The same is true for the Dirac cones in $\alpha$-(BEDT-TTF)$_2$I$_3$.
This unusual property is understood by considering the chirality associated with
the hole band around $\Gamma$ point as shown in Fig.~\ref{fig_chirality}(c).
Although the energy dispersion around the $\Gamma$ point is quadratic,
and an energy gap lies between two relevant bands,
we can apply the above analysis in constructing the effective Hamiltonian:
In the quadratic dispersion case, the parameters ${\bf v}_{\rho}$ 
are linear functions with respect to $k_x$ or $k_y$ with higher order corrections.
Along the circle represented by $k=\sqrt{k_x^2+k_y^2}=0.1$, we find
${\cal H}_{\bf k} = a_{\bf k} \sigma_0 + b_{\bf k} \sigma_z + c_{\bf k} \sigma_x$,
with $a_{\bf k} = -0.508k_x^2-0.475k_y^2$,
$b_{\bf k} = -0.191 k_x^2 + 2.56 k_y^2$,
and $c_{\bf k} = 2.80 k_x k_y$.
The winding is described by the following vector:
$\left( {n_x ,n_z } \right) = 
{\left( {c_k , b_k } \right)}/\sqrt{ b_k^2 + c_k^2}$.
As shown in Fig.~\ref{fig_chirality}(c), the vector ${\bf n}_{\bf k}$ 
rotates counter-clockwise twice when one goes around the $\Gamma$ point.
This chirality exactly compensates chirality of the two Dirac cones.
We note that the scattering between the two Dirac cones,
which is still under debate in graphene \cite{Jiang07},
is qualitatively different from that in graphene because these Dirac cones
carry the same chirality.
It is worth of pointing out that this chirality property
is compatible with the interband Cooper pairing between
the hole band around the $\Gamma$ point and the Dirac cones.

Now we turn to the physical properties associated with the Dirac fermions.
One of the most significant effects associated with the Dirac fermion chirality
is the suppression of backward scattering \cite{Ando98}.
As a consequence the scattering ratio is different
between Dirac fermions and conventional electrons.
Reflecting this difference
some transport coefficients clearly exhibit contributions 
from Dirac fermions
\cite{Fukuyama08}
even if the Dirac fermions are the minor carrier \cite{Harrison09}.

\begin{figure}[t]
   \begin{center}
    \includegraphics[width=0.7 \linewidth]{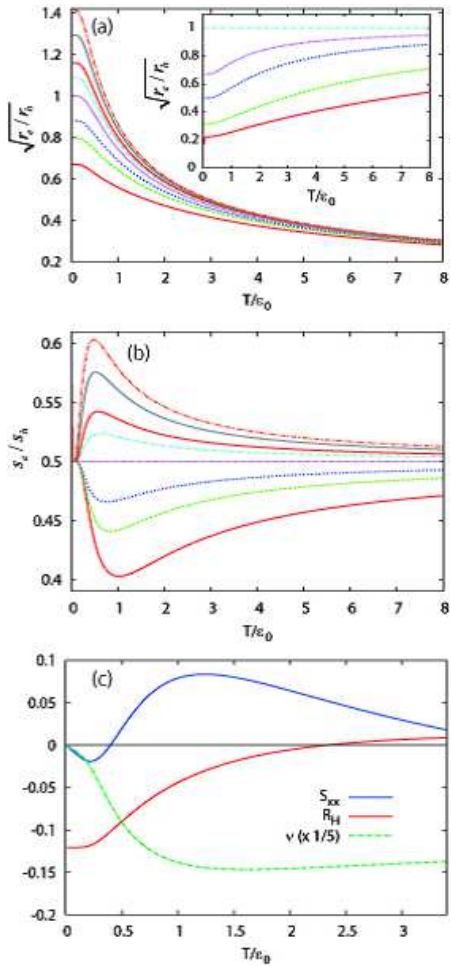}    
   \end{center}
   \caption{ 
   (color online)
    \label{fig_transport}
    The temperature dependence of the ratios $\sqrt{r_e/r_h}$ (a)
    and $s_e/s_h$ (b) for different values of the density ratio $n_r$.
    ($n_r = 1.0, 0.70, 0.45, 0.35, 0.25, 0.15, 0.10, 0.05$
	from the top to the bottom.)
    The inset in (a) shows the temperature dependence of
    $\sqrt{r_e/r_h}$ for conventional energy dispersions.
	($n_r = 1.0,0.45,0.25,0.10,0.05$ from the top to the bottom.)
    (c) The temperature dependence of 
    the Hall coefficient,
    the Seebeck coefficient,
    and the Nernst coefficient, 
    which are normalized by 
	$(2\pi a^2/|e|)(\hbar v/\varepsilon_0 a)^2$, $k_B/|e|$, 
        and $(k_B a^2/(2\pi \hbar))(\hbar v/\varepsilon_0 a)(v\tau_D/a)$, 
        respectively, with $k_B$ the Boltzmann constant.
    }
 \end{figure}

We consider a phenomenological two-band model 
consisting of a hole band, with a conventional energy spectrum,
and an electron band, with the Dirac fermion energy spectrum.
The charge current is given by
${\bf{j}} = \sigma {\bf{E}} 
+ \alpha \left( { - \nabla T} \right)$,
where $\sigma$ and $\alpha$ are tensors 
computed by the Jones-Zener solution 
of the Boltzmann equation \cite{ZimanEP}.
The contribution from the Dirac fermions is denoted by $\sigma^{(e)}$
and $\alpha^{(e)}$
and that from the holes is denoted by $\sigma^{(h)}$
and $\alpha^{(h)}$.
The sign change of the Hall coefficient
$R_H = \sigma_{xy}/(B_z \sigma_{xx}^2)$
with $B_z$ the applied magnetic field in the $z$ direction
occurs at $\sigma_{xy}^{(e)}/\sigma_{xy}^{(h)}=1$.
On the other hand, the sign change in the thermopower
occurs when $\alpha_{xx}^{(e)}/\alpha_{xx}^{(h)}=1$.
For the energy dispersions, we take
$\varepsilon _k^{\left( e \right)}  = vk - \varepsilon_e$
and $\varepsilon _k^{\left( h \right)}  =  
- k^2/(2m) + \varepsilon_h$
with energy levels $\epsilon_e$ and $\epsilon_h$.
At finite temperatures, electrons may be scattered by phonons and spin waves.
Here we focus on temperatures lower than the characteristic temperatures 
associated with those excitations and assume 
constant scattering times, $\tau_{e,h}$.
In order to reduce the number of parameters, 
we set $mv^2/2=\varepsilon_e \equiv \varepsilon_0$.
The difference between $\tau_e$ and $\tau_h$ is parameterized by
$\tau_r \equiv \tau_h/\tau_e$.
In Fig.~\ref{fig_transport}(a) and (b) we show 
$\sqrt{r_e/r_h} \equiv \sqrt{\tau_h^2 \sigma_{xy}^{(e)}/(\tau_e^2 \sigma_{xy}^{(h)})}$ 
and $s_e/s_h \equiv \tau_h \alpha_{xx}^{(e)}/(\tau_e \alpha_{xx}^{(h)})$
for differnt values of $n_r \equiv n_e/n_h$ with 
$n_{e(h)}$ the electron (hole) density.
Whether there are sign changes or not depends on $\tau_r$ and $n_r$.
(Note that $n_r$ is controlled by changing $\varepsilon_h$.)
The sign change occurs when $\tau_r = \sqrt{r_e/r_h}$, $s_e/s_h$.
In general, the sign change occurs at different temperatures 
for the Hall coefficient and the thermopower
because of their different dependences on $\tau_r$.

We show the Seebeck coefficient, the Hall coefficient,
and the Nernst coefficient, 
$\nu=(\alpha_{xx} \sigma_{xy} - \alpha_{xy} \sigma_{xx})
/\left[ B_z (\sigma_{xx}^2+\sigma_{xy}^2) \right]$,
in Fig.~\ref{fig_transport}(c) for $\tau_r=0.45$ and $n_r=0.05$.
The result is consistent with 
the experiment reported in Ref.~\cite{Matusiak10},
where the similar sign changes were observed in CaFe$_2$As$_2$.
Note that we expect large contributions from the Dirac fermions
for the Nernst coefficient as well because the quantity is quadratically dependent on
the scattering time.
We point out that a naive application of the Jones-Zener solution 
to the SDW mean field state with the same scattering time for all bands
does not lead to these temperature dependence
even if we take into account the temperature dependence of the order parameter.
In CaFe$_2$As$_2$, the energy dispersion in the $k_z$ direction may not be 
negligible.
Therefore, we assumed here a moderate value for $\tau_r$.
For systems with strong two-dimensionality,
like the so-called 1111 system, we expect a small $\tau_r$.
In such a case Dirac fermions markedly contribute
to the transport coefficients.

If we assume a conventional energy dispersion
for the electron band, the temperature dependence of 
$\sqrt{r_e/r_h}$ is qualitatively different from the Dirac fermion case
as shown in the inset in Fig.~\ref{fig_transport}(a).
In this case, the Hall coefficient behaves qualitatively different
from Fig.~\ref{fig_transport}(c).
The Hall coefficient is {\it positive} at low temperatures,
unlike the Dirac fermion case.

\begin{figure}[t]
   \begin{center}
    \includegraphics[width = 0.8 \linewidth]{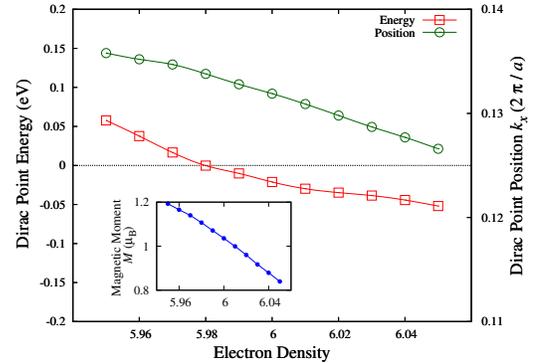}        
   \end{center}
   \caption{ 
   \label{fig_ndep}
   (color online)
	The electron density dependence of the Dirac point and
	the Dirac point position in the Brillouin zone denoted by
	$(k_x,0)$. 
	The inset shows the electron density dependence of 
	the magnetic moment.
    }
 \end{figure}

A systematic change of these sign changes 
are expected by controlling the electron density.
In Fig.~\ref{fig_ndep} we show the doping dependence of the Dirac point
and its position in the Brillouin zone
in the five-band SDW state.
Both decrease monotonically by increasing electron density.
Interestingly the Dirac point crosses the Fermi energy at the hole doping side.

Now we propose an experiment to directly observe the evidence of the Dirac fermions.
In conventional electrons, the Landau level energies are equally separated.
By contrast, the Dirac fermion Landau level energies are not.
For the tilted Dirac cone (\ref{eq_DiracH}), 
the Landau level energy spectrum is given by \cite{Morinari09}
\be
E_n = {\rm sgn}(n) \frac{\sqrt{{v_z^x} {v_x^y}}}{\ell_B} \sqrt{2 \lambda^3 |n|},
\ee
with $\ell_B=1/\sqrt{eB}$ the magnetic length and 
$\lambda = \sqrt{1-(v_0^x/v_z^x)^2}$.
This Dirac fermion Landau level structure 
can be observed by scanning tunneling spectroscopy (STS) measurements.
At graphite surfaces under magnetic fields
a series of peaks in the tunnel spectra associated 
with Landau quantization of the quasi-two-dimensional electrons were
observed \cite{Matsui05,LiAndrei07}.
The observed Landau levels are in good agreement with 
those expected for a surface layer, graphene.
From our effective Hamiltonian, we evaluated that
$E_1 \simeq 10$ meV at $B=10$ T while $E_1 \simeq 130$ meV in graphene.
Using STS, we expect that the Dirac fermion Landau level structure 
should be observed in the parent antiferromagnetic state.

To conclude, we have investigated the Dirac fermions
in the parent antiferromagnetic state of the iron-based superconductors.
There exist two tilted Dirac cones with the same chirality
contrary to the Dirac cones in graphene. 
The sign changes observed experimentally 
in the thermopower and the Hall coefficient
are consistently understood from the Dirac fermion picture.
We propose that the Dirac fermion Landau level structure 
be observed by STS measurements.
Since the two Dirac cones carry the same chirality, 
the intervalley scattering is qualitatively different from 
ordinary Dirac fermion systems. 
Compared to graphene, the valley splitting occurs in a different manner 
that would be clarified in the magnetoresistance.

This work was supported by the Grant-in-Aid for Scientific Research from the Ministry 
of Education, Culture, Sports, Science and Technology of Japan; 
the Global COE Program ``The Next Generation of Physics, Spun from University and Emergence"; 
and Yukawa Institutional Program for Quark-Hadron Science at YITP.
Numerical computation in this work was carried out at the Yukawa Institute Computer Facility.
E.K. is the Yukawa Fellow and this work was partially supported by Yukawa Memorial Foundation.

\bibliography{../../../references/tm_library2}

\end{document}